\documentclass[submission,copyright,creativecommons]{eptcs}
\providecommand{\event}{ICLP DC 2021}

\usepackage[utf8]{inputenc}
\usepackage{amsmath}
\usepackage{amssymb}
\usepackage{microtype}
\usepackage{url}

\usepackage{breakurl}             
\usepackage{underscore}           

\newcommand{\sysfont}{\textit}

\newcommand{\Gringo}{\sysfont{Gringo}}

\newcommand{\clasp}{\sysfont{clasp}}

\input{dtel}

\def\titlerunning{Formalisation of Action with Durations in ASP}
\def\authorrunning{E. Tignon}

\begin{document}

\title{Formalisation of Action with Durations in Answer Set Programming}

\author{%
  Etienne Tignon\\
  University of Potsdam, Germany
}

\maketitle

\begin{abstract}
In this paper, I will discuss the work I am currently doing as a Ph.D. student at the University of Potsdam, under the tutoring of T. Schaub.
I’m currently looking into action description in ASP.
More precisely, my goal is to explore how to represent actions with durations in ASP, in different contexts.
Right now, I’m focused on Multi-Agent Path Finding (MAPF), looking at how to represent speeds for different agents and contexts.

Before tackling duration, I wanted to explore and compare different representations of action taking in ASP.
For this, I started comparing different simple encodings tackling the MAPF problem.
Even in simple code, choices and assumptions have been made in their creations.
The objective of my work is to present the consequences of those design decisions in terms of performance and knowledge representation.
As far as I know, there is no current research on this topic.

Besides that, I’m also exploring different ways to represent duration and to solve related problems.
I planed to compare them the same way I described before.
I also want this to help me find innovative and effective ways to solve problems with duration.
\end{abstract}

\section{Introduction}\label{sec:introduction}

ASP is already used in time-related solving in a lot of different contexts, from planning trains \cite{abjoossctowa19a} to moving robots in wharehouse \cite{ngobsoscye17a} \cite{goheba21a}.
There exist some dedicated tools like Telingo \cite{cakamosc19a} that are used for planning.
However, most of the existing works make an assumption: that the events you are trying to place in time are instantaneous (in the sense that they are defined in a point in time, opposed to an interval).
This is not always what you want to represent. For example: what if you want to create a schedule for the lectures in a university?
The lectures are not instantaneous, they take time.
How to represent things who take times?
The goal of my Ph.D. is to study how to represent timed events with duration in ASP.
I'm looking at different implementations of timed events in ASP, comparing them in terms of knowledge representation and performances.
From that, I hope to be able to deduce performant and easy-to-read models of timed events with duration.

For now, even if I hope my results to be versatile, I'm mostly focused on solving MAPF with duration.
In MAPF with duration, agents can take actions to move from one point to another like in classical MAPF.
But the time needed to do this action can change depending on several parameters.
But I would like to use ASP and its strengths to find innovative ways to represent and solve MAPF instances where different durations of actions have to be taken into consideration.

\section{Background}\label{sec:background}

\subsection{Answer Set Programming}

Answer Set Programming \cite{gellif88b} \cite{subzan95a} is a versatile declarative problem-solving paradigm.
It combines a high-level modeling language with effective grounding and solving technology.
On top of that, its success is also due to effective systems, like the solver $\clasp$ \cite{gekanesc07b} and the grounder $\Gringo$ \cite{gescth07a} from the Potsdam Answer Set Solving Collection (aka Potassco) \cite{gekakasc12a}.

Research has already been made that uses ASP to tackle time-related problems.
For exemple, ASP is explored to solve MAPF \cite{ngobsoscye17a} \cite{goheba21a} or train scheduling \cite{abjoossctowa19a}.
Also, there exist systems dedicated to this, like telingo \cite{cakamosc19a}.
ASP implementations of some of the mentioned representations exists (for action languages \cite{lifschitz99a},\cite{dwgrniscse07a},\cite{gegrsc10a}, event calculus \cite{kilepa09a}, PDDL+ \cite{bamama16b} to name a few).

\subsection{Multi-Agent Path Finding}

As I said before, during this Ph.D., I am mostly interested in applying techniques and concepts of the formalisms on MAPF \cite{ststfekomawaliatcokubabo19b}.
The Multi-Agent Path Finding problem is an extension of the pathfinding one for which the path of multiple agents has to be computed while taking into consideration the interactions and possible collisions among the agents.
Most of the current work on MAPF defines a discrete environment where the space to explore is defined by nodes on which the agent can be at any point in time.
The edges between two nodes represent the fact that an agent can move from one to another.
In this context, a move doesn't really take time.
This implies that if at time $t$ an agent is on the node $n_1$, if it moved in the edge $n_1,n_2$, at time $t+1$ it will be on the node $n_2$.
What I am interested in during this Ph.D. is to explore ways to represent the concept of ``this agent takes $X$ time to move from $n_1$ to $n_2$".

Several works deal with similar subjects, like with continuous time \cite{anyaatst19a}, or with geometric agents \cite{surynek19b}.
However, most of these works are not related to ASP.
Moreover, I want to study actions with duration in particular.
Most of the literature about MAPF when movements take time also makes other assumptions, such as space and/or time being continuous for example.

\subsection{Timed Events Formalisms}

The will to express timed events rigorously and practically is far from new in the field of knowledge representation.
Here is a brief, non-exhaustive list of formalizations and models existing in the literature which are planned to be explored in this Ph.D.
\begin{itemize}
  \item \textbf{Action languages} \cite{pednault87a}\cite{turner97a}\cite{giulif98a}\cite{gellif98a} are formal models used for talking about actions and their effects. They use a transition system composed of fluents which have different values that can be modified by actions. Depending on the exact language used, different advanced concepts can be expressed, like conditional consequences on an action, redefinition of inertia, action with indirect effects, and many others.
  \item \textbf{Allen's interval algebra} \cite{allen83a} is a calculus made for temporal reasoning. It allows the description of relationships between intervales. Thanks to this formalizations, it is possible to represent the relations between timed events (such as ``the event $e_1$ occur before the event $e_2$''). This calculus doesn't represent any quantitative notion like the time of the intervals or the distance between two intervals in time; only qualitative relations are exprimed.
  \item \textbf{Event calculus} \cite{kowser86a}, like action languages, is a formalism able to represent events and their effects. In this formalism, you have fluents which can either hold or not at each point in time. Furthermore, you have events which can occur and change fluent which holds. You can, for example, express that an action $\alpha$ makes a fluent $\beta$ hold at time $\tau$ under the condition that another fluent $\beta'$ hold at the same time.
  \item \textbf{PDDL+} \cite{foxlon06a} is a moddeling language destined to represent mixed discrete-continuous planning domains. It is an extention of \textbf{PDDL} \cite{mcdermott98a}. Inside this formalism, it is possible to represent events and actions which have effect on continuous variables.
  \item \textbf{Temporal logic} \cite{gabbay87a} \cite{abaman89a} \cite{degola16a} is a way to extend logic with temporal symbols, such as next ($\circ$), eventually ($\diamond$), or until ($\until$). In this paradigm, you consider that atoms are true or false at each point in time. With thoses temporal symbols, you can write the relationships between different timepoints. For example to say that, if the atom $a_1$ holds, $a_2$ hold in the next timepoint, you can write $a_1 \rightarrow \circ a_2$.
  \item \textbf{Metric temporal logic} \cite{koymans90a} further extends temporal logic by using time-constrained version of the temporal operators. With this, it is possible to represent quantified timed relationships, like ``if the atom $a_1$ hold, $a_2$ hold until in 10 $a_3$ hold (wich mean that $a_3$ must hold at least once in the next 10 timepoints) :  $a_1 \rightarrow a_2 \, \until_{10} \, a_3$"
  \item \textbf{Temporal network} \cite{tn1} is a formalism that allow the representation of graphs that have edges defined for each point in time (for example, you can have the edge $(n_1,n_3,4)$ to represent the edge from $n_1$ to $n_3$ at time $4$).
\end{itemize}

\section{Current status}\label{sec:currentstatus}

For now, most of the work has been dedicated to the study of actions representations, durations, MAPF, and ASP.
A solid understanding of the state of the art in those fields is needed to ensure that the work of this Ph.D. will lead to a significant advance.

Besides that, I have started a comparative analysis of different encodings based on action languages in ASP, dedicated to the solving of MAPF instances.
In particular, two aspects are observed:
\begin{itemize}
  \item{Knowledge representation}: Some implementations are easier to read than others. It is important to not underestimate the importance of it, so we can create understandable and maintainable programs.
  \item{Performance}: One line of difference can have a great impact on the behavior of the grounder and the solver. Most of the time, the same concept can be modeled in many ways. Each of them will lead to a different number of atoms and rules, which will define how long the grounder will need to create the grounded program. This, in turn, will impact how the solving will be tackled.
\end{itemize}

I hope this structured comparison between encodings will lead me to a better understanding of the best way to model this kind of problem in ASP.

I also started thinking about how to represent MAPF problems in the differents paradigms discussed earlier.
I intend to have the same example in all those formalizations so that in the future, when I will look at their implementations in ASP, I will have a common ground to comparison.

\bibliographystyle{eptcs}
\bibliography{krr,procs}

\end{document}